\documentclass{article}

\usepackage{arxiv}

\usepackage[utf8]{inputenc} 
\usepackage[T1]{fontenc}    
\usepackage{hyperref}       
\usepackage{cancel}
\usepackage{url}            
\usepackage{booktabs}       
\usepackage{amsfonts}       
\usepackage{nicefrac}       
\usepackage{microtype}      
\usepackage{lipsum}		
\usepackage{graphicx}
\usepackage{soul}
\usepackage{doi}
\usepackage{xcolor}
\usepackage{amsmath}
\usepackage{todonotes}

\title{The Dependence of Parallel Imaging with Linear Predictability on the Undersampling Direction}

\author{ Alex McManus \\
	Department of Applied Mathematics\\
	University of Colorado Boulder\\
	Boulder, CO \\
	\And
	Stephen Becker\\
	Department of Applied Mathematics\\
	University of Colorado Boulder\\
	Boulder, CO \\
	\And
	Nicholas Dwork\\
	Department of Biomedical Informatics\\
	University of Colorado School of Medicine\\
	Aurora, CO \\
}

\renewcommand{\shorttitle}{Directionality in Parallel MRI}

\begin{document}
\maketitle

\begin{abstract}

Parallel imaging with linear predictability takes advantage of information present in multiple receive coils to accurately reconstruct the image with fewer samples. Commonly used algorithms based on linear predictability include GRAPPA and SPIRiT. We present a sufficient condition for reconstruction based on the direction of undersampling and the arrangement of the sensing coils. This condition is justified theoretically and examples are shown using real data. We also propose a metric based on the fully-sampled auto-calibration region which can show which direction(s) of undersampling will allow for a good quality image reconstruction.

\end{abstract}

\keywords{Parallel Imaging \and Linear Predictability \and Fourier Shift Theorem \and SMASH \and GRAPPA \and SPIRiT}

\section{Introduction}

Magnetic resonance imaging (MRI) is a ubiquitously used and highly versatile gross medical imaging modality with abundant natural contrast; it is relatively slow due to the length of time required for longitudinal magnetization recovery, parameterized by $T_1$.  Producing images of diagnostic quality requires patients to remain still for a long time.  For a given sequence and contrast, scan time  is directly proportional to the amount of data collected---therefore, it is advantageous to reconstruct good quality images with fewer samples. 

Several methods to accelerate MRI have been developed, including parallel imaging \cite{deshmane2012parallel}, compressed sensing (CS) \cite{lustig2007sparse,baron2018rapid,dwork2021utilizing,dwork2022utilizing}, partial Fourier acquisition \cite{noll1991homodyne}, and, more recently, the use of convolutional neural networks and other deep learning solutions \cite{sandino2020compressed,hammernik2019sigma,knoll2020advancing}.

Parallel imaging uses multiple sensing coils to reduce the amount of data required for a high-quality image.  There are two distinct approaches to parallel imaging: model based reconstruction (which requires estimates of the sensitivity maps), and linear predictability (which uses a calibration region to estimate linear interpolation coefficients).  The quintessential algorithms for parallel imaging with linear predictability are AUTO-SMASH \cite{jakob1998auto}, GRAPPA \cite{griswold2002generalized}, and SPIRiT \cite{lustig2010spirit}.  In \cite{haldar_predictability}, Haldar et al. show that a sufficient condition for parallel imaging with linear predictability is that the support of the imaged object is a strict subset of the field-of-view of the image.  That is, if there are regions of the field-of-view that do not image the subject and just image air, then parallel imaging with linear predictability is possible.  Notably, though this is a sufficient condition for parallel imaging with linear predictability generally, it may not be sufficient for a specific instance of linear predictability based on user specified parameters.  Thus, even when this condition is met, parallel imaging algorithms may fail to yield a high-quality image, as we will show. 

In this manuscript, we will present a separate sufficient condition for parallel imaging with linear predictability. We will show that parallel imaging reconstruction algorithms based on linear predictability, such as GRAPPA \cite{griswold2002generalized} and SPIRiT \cite{lustig2010spirit}, have an innate dependence on the direction of undersampling, which is based on the arrangement of the sensing coils. We explore this dependence by showing that GRAPPA (and by extension, SPIRiT) is based on similar assumptions as those required by AUTO-SMASH \cite{jakob1998auto}. We show results where the quality of the reconstruction differs dramatically when the direction of undersampling is changed.  We also show cases where the quality is impervious to the direction of undersampling and explain why this is the case.  Lastly, we propose a metric based on the fully sampled auto-calibration region; this metric identifies which undersampling direction (horizontal or vertical) will lead to a low-quality image.

For simplicity, the discussion of this manuscript will be restricted to Cartesian sampling; the extension to parallel imaging with linear predictability when sampling with a non-Cartesian trajectory is as expected based on prior work \cite{seiberlich2007non,wright2014non,luo2019grappa}.

\section{Theory}
\label{sec:Theory}
In this section, we start by reviewing the main points of AUTO-SMASH \cite{jakob1998auto} and show that the generalization of its theoretical basis justifies GRAPPA \cite{griswold2002generalized} and SPIRiT \cite{lustig2010spirit}.  Note that a complete review of the theory of AUTO-SMASH using the notation of this manuscript can be found in Appendix \ref{app:autosmash}.  For the purposes of this discussion, we will assume that imaging is performed in two-dimensions.  These sampling patterns may be generated with a spin-warp trajectory using two dimensions of phase encodes and one dimension of readout; after inverse Fourier transforming along the readout direction, the data is placed in a hybrid space and the reconstruction of each slice may be considered independent of every other slice \cite{beatty2007method}.

For this discussion, the MRI signal with spin density $\rho:\mathbb{R}^2\rightarrow[0,\infty)$ for coil $j$ is
\begin{equation}
  \label{eqn:mriSignal}
  S_{j}(k_x, k_y) = \iint dx \, dy \, C_j(x, y) \, \rho(x, y) \, \exp\left(-i\, k_x\, x -i\, k_y\, y\right)
                  = \mathcal{F}\left\{ C_j \, \rho \right\}\left(k_x,k_y\right),
\end{equation}
where $k_x = \gamma \int_0^{t_x} G_x(\tau) d\tau$, $k_x = \gamma \int_0^{t_y} G_y(\tau) d\tau$, $\gamma$ is the gyromagnetic ratio, $G_x$ and $G_y$ are the x and y gradient waveforms, $t_x$ and $t_y$ are the lengths of time that the respective gradient fields are turned on, and $\mathcal{F}\{\rho\}(k_x,k_y)$ denotes the Fourier transform of $\rho$ evaluated at $(k_x,k_y)$.
For simplicity, we are ignoring the effects of relaxation and recovery.

\begin{figure}[ht]
  \centering
  \includegraphics[width=0.5\linewidth]{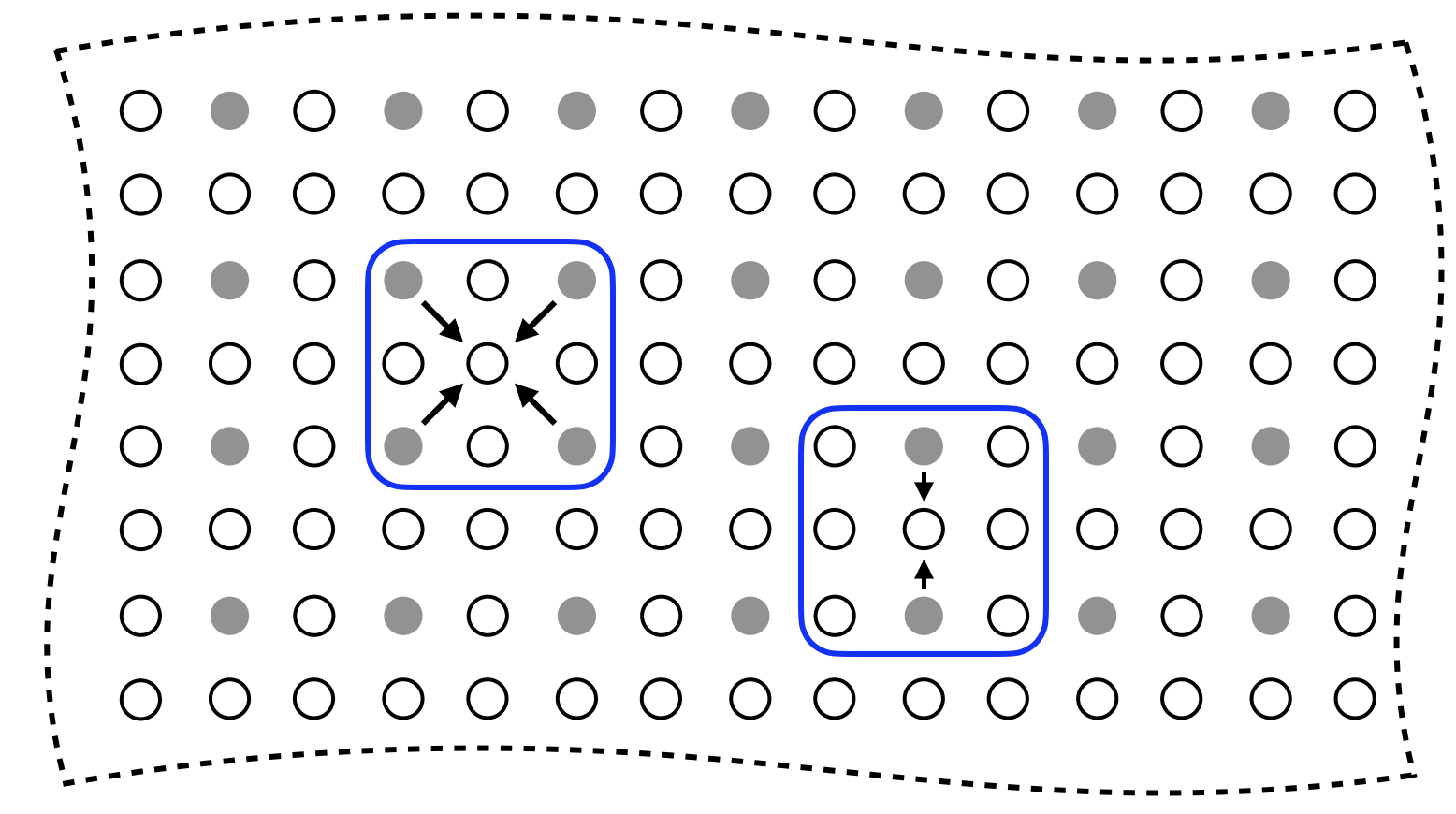}
  \caption{In this depiction of parallel imaging with linear prediction, a portion of a sampling pattern is shown: filled in circles represent data that was collected and unfilled circles represent data that was not.  A distance threshold of $1$ and a metric of $\|\cdot\|_\infty$ are depicted with blue contours. The undersampling rate in each direction leads to different patterns of collected and uncollected points.  The arrows represent displacements between collected data that lie within the threshold distance and a point an uncollected point that must be estimated.}
  \label{fig:grappaDisplacements}
\end{figure}

Parallel imaging with linear predictability is depicted in Fig. \ref{fig:grappaDisplacements}.  In this example, the data is undersampled by a factor of $2$ in both the horizontal and vertical directions (for a total undersampling factor of $4$).  Note that with $J$ coils, each sampled point consists of $J$ complex values, one for each coil.

The user supplies a metric and (at least one) threshold, perhaps one per direction.  Points that were not collected are synthesized as linear combinations of those values that lie within the threshold.  For the example provided in Fig. \ref{fig:grappaDisplacements}, the user has specified the $\|\cdot\|_\infty$ as the metric and a single threshold of $1$.  Thus, for any uncollected Fourier location, linear coefficient are found to interpolate from all collected points that lie within a threshold's distance.

\subsection{AUTO-SMASH}
AUTO-SMASH is a seminal work of parallel imaging with linear predictability; the algorithm uses a two-dimensional spin-warp trajectory \cite{epstein2007introduction} (e.g., phase encode and readout in $k_y$ and $k_x$ dimensions, respectively). Whereas the Nyquist sampling theorem would dictate that the phase encoding lines be separated by $\Delta k_y$, AUTO-SMASH separates lines by $M \Delta k_y$ for some $M>1$.  Consider the case where there are $J$ coils, each with sensitivity map $C_j$.  AUTO-SMASH defines a composite sensitivity map, $C_0^{\text{comp}}$, created by a linear combination of the individual coil sensitivity maps with linear coefficients $n^{(0)}\in\mathbb{C}^L$. That is, $C_0^{\text{comp}} = \sum_j n_j^{(0)} \, C_j$.
AUTO-SMASH further assumes that for each $m=1,\ldots,M-1$; there exist unknown linear coefficients $n^{(m)}\in\mathbb{C}^L$ such that $\sum_j n_j^{(m)} \, C_j \approx C_0^{\text{comp}} \, \exp(i\,m\,\Delta k_y\,y)$.
Then, by the Fourier shift theorem (as detailed in Appendix \ref{app:autosmash}), $\sum_j n_j^{(m)} \, S_j(k_x,k_y) = \mathcal{F}\left\{C_0^{\text{comp}}\rho\right\}(k_x,k_y-m\Delta k_y)$.
In \cite{epstein2007introduction}, the coils were designed to satisfy the above requirements.

This illustrates that the linear coefficients $n^{(m)}$ permit interpolation from known values to unknown values located $m\Delta k_y$ distance away in the $k_y$ direction of the Fourier domain.  It is assumed that these same coefficients are valid across the entire Fourier domain.
With AUTO-SMASH, the coils were designed so that $C_0^{\text{comp}}\approx \kappa$, for some constant $\kappa$, across the field-of-view of the image.
To determine the linear coefficients $n^{(m)}$, a set of lines in the Fourier domain (called the auto-calibration signal -- ACS) are collected and a linear system $\mathcal{S}\,n^{(m)}=s_{\text{acs}}$ is numerically solved, where $s_{\text{acs}}$ is comprised of the values of the ACS. (See Appendix \ref{app:autosmash} for details.)

\subsection{GRAPPA}
GRAPPA was developed heuristically; and yet, we have found that it can be explained with a theoretical basis very similar to that of AUTO-SMASH, as we will now do.

GRAPPA eliminates the assumption of known $n^{(0)}$ and the existence of an approximately constant composite sensitivity map.  
Moreover, instead of searching for coefficients that correspond to a single predetermined displacement, GRAPPA attempts to find linear coefficients for multiple displacements. The combination of distance threshold and undersampling rate gives rise to different patterns of collected and uncollected points used for interpolation, as depicted in Fig. \ref{fig:grappaDisplacements}. Each unique pattern of collected points surrounding an uncollected point is called a kernel; The blue contours in Fig. \ref{fig:grappaDisplacements} shows two different kernels.

For a given kernel, as with AUTO-SMASH, the linear interpolation coefficients can be determined by solving a linear system.  Here, we present a least-squares problem that simultaneously identifies the interpolation coefficients for all coils:
\begin{equation}
  \label{eqn:grappa}
  \underset{N}{\text{minimize}} \hspace{0.5em} \lVert \mathcal{S}\,N - s_{\text{acr}}\rVert_2^2,
\end{equation}
where $\mathcal{S}$ is comprised of the appropriate values from the auto-calibration region, $N$ is the matrix of weights we solve for, and $s_{\text{acr}}$ is a matrix of points from the auto-calibration region, a region which has rectangular size $n_x\times n_y$.
For a specific uncollected Fourier location $k$, let $D_k$ be the number of nonzero points of the relevant kernel.
The matrix $\mathcal{S}$ will be of size $\left[\eta_x\,\eta_y \, \times \, J\,D_k\right]$, $N$ is a matrix of size $\left[J\,D_k \, \times \, J\right]$, $s_{\text{acr}}$ is a matrix of size $\left[\eta_x\,\eta_y \, \times \, J\right]$, and $\eta_x$ and $\eta_y$ are the number of times that the kernel fits inside the auto-calibration region in the $k_x$ and $k_y$ directions, respectively.


We will now explicitly relate problem \eqref{eqn:grappa} to the assumption of Eq. \eqref{eq:grappaAssumption}.  An equivalent form of problem \eqref{eqn:grappa} is
\begin{equation}
  \label{eqn:grappa2}
  \underset{N}{\text{minimize}} \hspace{0.5em} \sum_{\mathbf{k}}\,\lvert \mathcal{S}^{(k)}N - s_{\text{acr}}^{(k)}\rvert^2,
\end{equation}
where $k\in\mathbf{k}$ is an individual location inside the auto-calibration region,  $\mathcal{S}^{(k)}$ is the row of $\mathcal{S}$ that corresponds to location $k$, and $s_{\text{acr}}^{(k)}$ is the $k^{\text{th}}$ row of $s_{\text{acr}}$. We recognize further that we can write $\mathcal{S}^{(k)}\,N$ as

\begin{equation}
  \label{eqn:grappa_inner}
  \mathcal{S}^{(k)}\,N = \sum_{d\in \mathcal{K}_k} \sum_{j} S_j(k + d) \, n_j^d
\end{equation}
where $S_j(k+d)$ is the signal collected at point $k+d$ from coil $j$, $n_j^d$ is the appropriate weight, the inner sum is over the coils, and the outer sum is over the collected sample points that lie within the kernel $\mathcal{K}_k$.

Similarly to AUTO-SMASH, for a given displacement $d=(u\,\Delta k_x,v\,\Delta k_y)$ and coil $\ell$, GRAPPA seeks a set of linear coefficients $n^{(d)}\in\mathbb{C}^J$ such that
\begin{equation}
  \label{eq:grappaAssumption}
  \sum_j n_j^{(d)} \, C_j \approx \exp\left(i\left( u\Delta k_x \, x + v\Delta k_y \, y\right) \right) C_\ell.
\end{equation}
The notable difference is that the linear combination yields a complex exponential weighted by an individual coil's sensitivity rather than a composite sensitivity.  Using analogous mathematics as presented in App. \ref{app:autosmash}, performing a linear combination of the collected points from all coils linearly interpolates missing values of a specific coil's data.  After the missing data of all coils is interpolated, the data from multiple coils can be combined into a single image \cite{roemer1990nmr}.

Here, we describe a sufficient condition for estimation without error when the approximation of \ref{eq:grappaAssumption} is perfectly satisfied and without any noise.  With the formulation described above, the sufficient condition becomes evident.  This formulation shows that each point of the GRAPPA kernel corresponds to a set of weights (one for each coil). Each interpolated point can be thought of as a linear combination of sensitivity maps approximating a complex exponential of a specific frequency in accordance with Eq. \eqref{eq:grappaAssumption}. \textbf{\textit{For a given location $k$ and kernel, if at least one displacement vector to a collected point within the kernel satisfies Eq. \eqref{eq:grappaAssumption}, then the interpolation will be accurate.}}  If more than one displacement vector satisfies Eq. \eqref{eq:grappaAssumption}, then GRAPPA finds the set of linear coefficients that interpolate from multiple points in a least-squares optimal sense.

\subsection{SPIRiT}
SPIRiT \cite{lustig2010spirit} is an extension of GRAPPA. With a fixed kernel size, rather than just interpolating from points that were collected, SPIRiT will use \textit{every point} in the kernel, regardless of whether or not it was collected.

SPIRiT interpolates all values (even those that were collected) by solving the following constrained least-squares problem:
\begin{equation}
  \label{eqn:spr2}
    \underset{\theta}{\text{minimize}} \hspace{0.5em}\frac{1}{2}\lVert G\theta - \theta \rVert_2^2 \hspace{0.5em}
    \text{subject to}\hspace{0.5em} \lVert{D\theta - y}\rVert_2^2 \leq \varepsilon,
\end{equation}
where $\|\cdot\|_2$ denotes the $l_2$ norm, $G$ represents linear interpolation from all values that lie within the kernel (even those that were not collected), $D$ is the linear transformation that isolates the sample points that were collected and, $y$ is the values of the collected data, and $\varepsilon$ is a bound on the noise power.  In general, problem \eqref{eqn:spr2} can be solved with the Fast Iterative Shrinkage-Thresholding algorithm (FISTA) \cite{beck2009fast}.  In \cite{lustig2010spirit}, Lustig et al. set $\varepsilon=0$ and only solve for the values of the uncollected data, which somewhat simplifies the implementation of the optimization algorithm, but this is not necessary.

Here, again, we describe a sufficient condition for estimation without error when the approximation of Eq. \ref{eq:grappaAssumption} is perfectly satisfied and without any noise.  The sufficient condition for an accurate SPIRiT interpolation is similar to that which was developed for GRAPPA, but more general. 
Consider the set of collected data as a directed graph where the location of each Fourier value is a node and the displacement vectors from each point in the kernel to that location are the directed edges.  \textbf{\textit{Accurate interpolation at a location $k$ is possible when there is a path from a collected data point to $k$ such that all edges of that path satisfy the approximation of Eq. \eqref{eq:grappaAssumption} well.}} Practically, any error in the approximation and noise in the values are amplified with each edge of the path; so shorter paths lead to more accurate interpolations.

\begin{figure}[ht]
  \centering
  \includegraphics[width=0.4\linewidth]{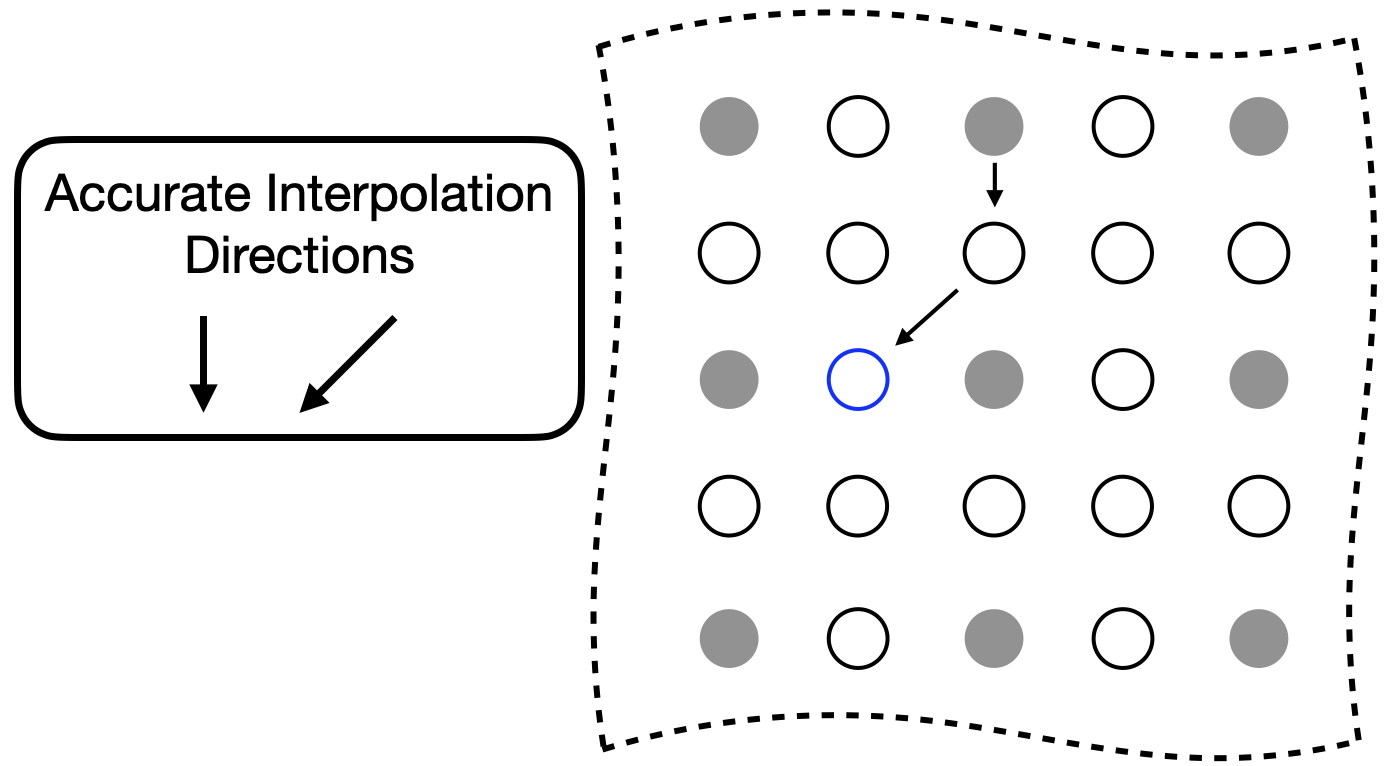}
  \caption{ A depiction of accurate interpolation with SPIRiT.  For this hypothetical example, the directions that the approximation of Eq. \ref{eq:grappaAssumption} is satisfied are down and downward-left.  Without noise, the value of the blue circle can be estimated by two interpolations along the path indicated with the arrows. }
  \label{fig:spiritSufficient}
\end{figure}

Figure \ref{fig:spiritSufficient} depicts an example that explains this sufficient condition.  In this example, we would like to interpolate the value of the blue circle.  The directions where condition Eq. \ref{eq:grappaAssumption} is satisfied is down and downward-left.  We could interpolate the value of the blue circle in two steps: 1) estimate all values below all sampled values, and 2) estimating all values downward and left of all known or estimated values.  This would perfectly estimate the value of the blue circle.  If this process were iterated until all unknown values were estimated, then it would solve problem Eq. \ref{eqn:spr2} with an objective function value of 0.

\subsection{Accuracy Metric}
\label{sec:metric}
In the results section, we will provide examples where the quality of the reconstruction is significantly degraded when the sufficient conditions identified are not met.  Here, we present a metric that can be calculated from the ACR that identifies any undersampling directions that would not meet the sufficient conditions for accurate interpolation. 
Given the fully sampled ACR and a kernel size, we solve for the interpolation coefficients using two kernels---one solely in the horizontal direction and one in the vertical direction---of the same size as the kernel used for reconstruction.  The kernel has a $0$ in the center and all other values are $1$.  E.g., if the desired kernel is $3\times3$, the test kernels look like
\begin{equation}
  \label{eqn:kernels}
  \mathbf{k}_h = \begin{bmatrix} 1 &0 &1\end{bmatrix}, \qquad \mathbf{k}_v = \begin{bmatrix}1\\0\\1\end{bmatrix}.
\end{equation}
Denote the solutions to problem \eqref{eqn:grappa} with each of these kernels as $N^\star_h$ and $N^\star_v$.
The metric for accuracy is the relative norm error of the linear system:
\begin{equation}
  \lVert \mathcal{S}\,N^\star - s_{acr}\rVert / \lVert{s_{acr}}\rVert
\end{equation}

If the value of the relative error is high, then there is not a consistent set of interpolation coefficients for the auto-calibration region in the direction specified by the kernel.  If the value of the relative error is low, then there is and the data can be undersampled in that direction while retaining a high-quality reconstruction.

\section{Experimental Setup}
In this manuscript, we first use Biot-Savart simulations to examine an 8-element birdcage coil.  We then analyze three different datasets: a knee, a brain, and an ankle.  All datasets were fully-sampled three-dimensional Cartesian data with two dimensions of phase encodes and one dimension of readout.  The data of the knee was taken from the publicly available MRIData \cite{ong2018mridata}.  Data of the brain and ankle were collected with a 3 Tesla General Electric MR750 clinical scanner (GE Healthcare, Waukesha, WI).  All procedures performed in studies involving human participants were in accordance with the ethical standards of the institutional and/or national research committee and with the 1964 Helsinki Declaration and its later amendments or comparable ethical standards. MR data of humans were gathered with institutional review board (IRB) approval and Health Insurance Portability and Accountability Act (HIPAA) compliance. Informed consent was obtained from all individual participants included in the study.

Data was retrospectively downsampled for processing.  The data was inverse transformed along the readout direction and placed into a hybrid space of $(k_x,k_y,z)$.  Then, individual slices at specific $z$ locations were isolated for further processing.

\section{Results}
\label{sec:results}

\begin{figure}[ht]
  \centering
  \includegraphics[width=6in]{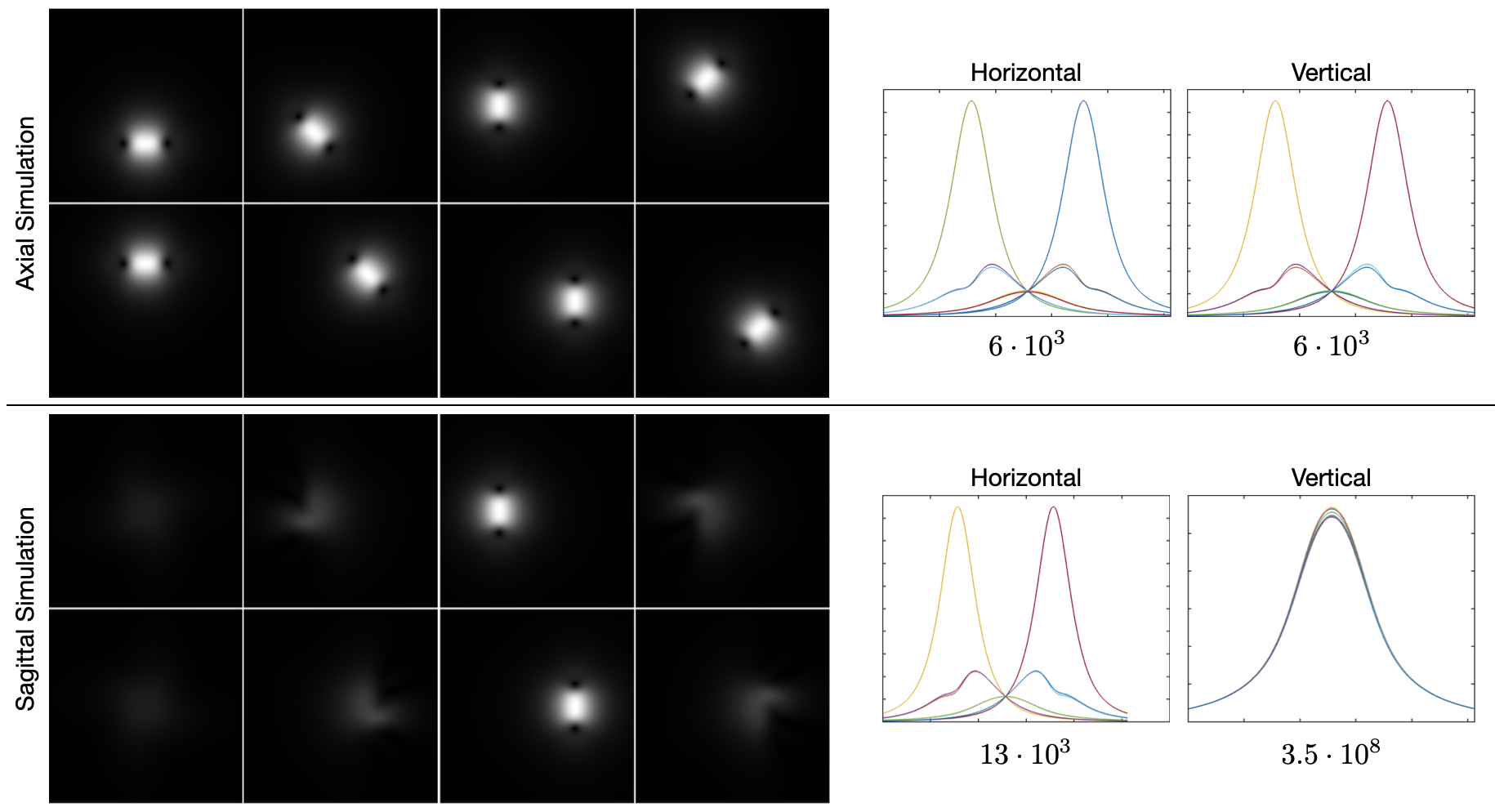}
  \caption{Biot-Savart simulations of an 8-element birdcage coil oriented in perpendicular directions.  (The top/bottom simulation would be the center axial/sagittal slice for a brain coil, respectively).  The sensitivity maps of each coil are shown on the left.  The plots on the right show the sensitivities of each coil for a single horizontal/vertical line through the center of the sensitivity maps.  The numbers below each plot show the condition number of a matrix created by concatenating the sensitivities in the plots above.}
  \label{fig:coilProfiles}
\end{figure}

Figure \ref{fig:coilProfiles} shows results from Biot-Savart simulations for two perpendicular orientations of an 8-element birdcage coil according to \cite{giovannetti2002fast}.  The top/bottom of the figure shows simulations for an axial/sagittal plane that lies at the birdcage coil, respectively.  The plots isolate single horizontal and vertical lines that lie at the center of the simulations; the number below each plot is the condition number of the matrix made by concatenating the vectors of the plots above it.  This condition number is a metric that indicates how much variation there is between the sensitivity maps across space.  For both the horizontal and vertical lines of the axial simulations, the condition number is on the order of $10^3$.  While this remains the case for the horizontal line of the sagittal simulation, the condition number for the vertical line is much higher: on the order of $10^8$.  By looking at the corresponding plot, it becomes obvious that the sensitivities are approximately scaled versions of each other, and that the problem of finding coefficients to linearly combine the sensitivity maps so that they approximate a complex exponential is ill-conditioned. 

\begin{figure}[ht]
  \centering
  \includegraphics[width=1.5in]{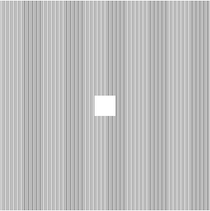} \includegraphics[width=1.5in]{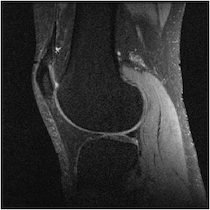}\,\includegraphics[width=1.5in]{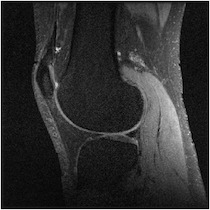} \\
  \includegraphics[width=1.5in]{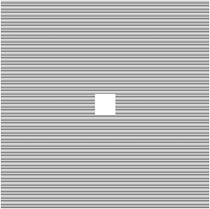}\,\includegraphics[width=1.5in]{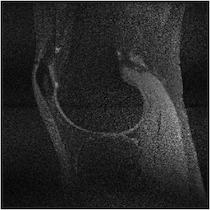} 
  \includegraphics[width=1.5in]{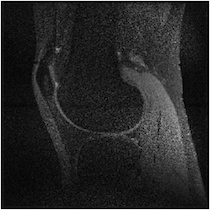}
  \caption{Reconstructions of sagittal slices of a knee with different sampling patterns.  The columns from left to right are the sampling pattern, the GRAPPA reconstruction, and the SPIRiT reconstruction.  The top and bottom rows show undersampling in the horizontal and vertical directions, respectively.  Both sampling masks used a reduction factor of 2.  All reconstructions used a $31\times 31$ ACR and a $3\times 3$ kernel.}
  \label{fig:fig1}
\end{figure}

Figure \ref{fig:fig1} shows reconstructions with GRAPPA and SPIRiT using two separate sampling patterns. Both data were retrospectively undersampled at the same reduction factor of 2; the only difference is the direction of undersampling. This dependence of quality on undersampling direction is present with both GRAPPA \cite{griswold2002generalized} and SPIRiT \cite{lustig2010spirit}.

\begin{figure}[ht]
  \centering
  \includegraphics[width=6in]{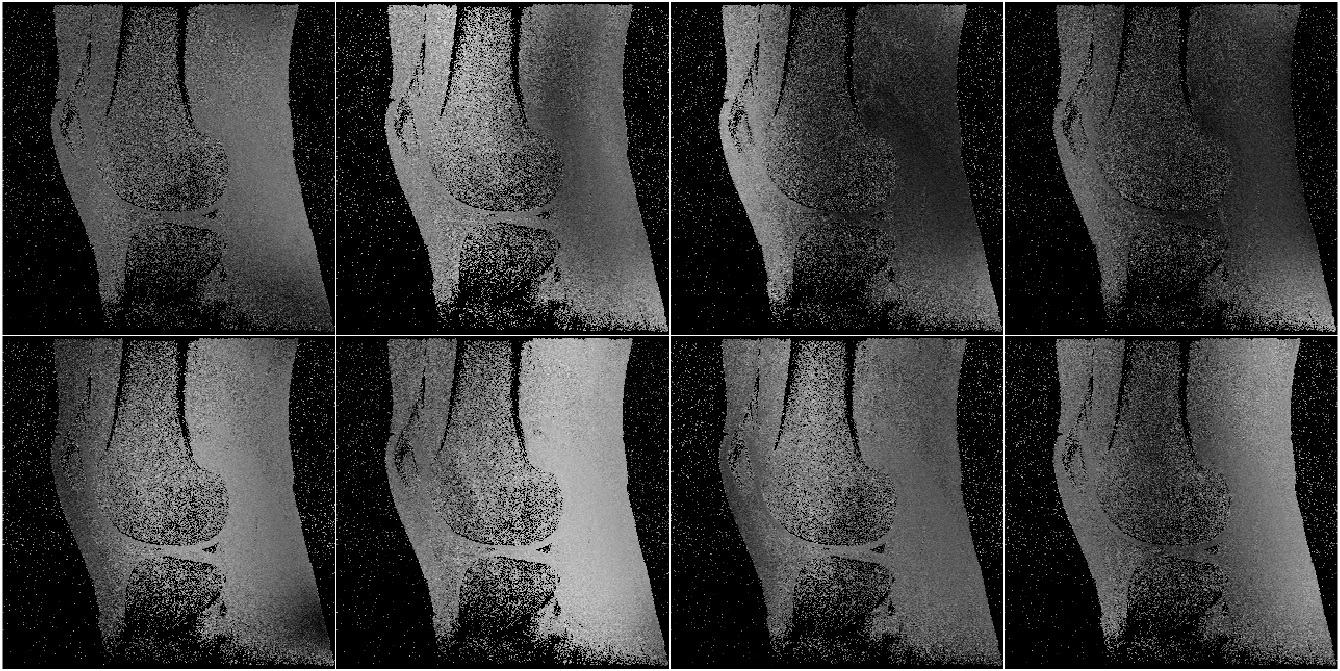}
  \caption{Approximations of the sensitivity maps for each coil in an 8-coil birdcage for the data of Fig. \ref{fig:fig1}.  Note that there are only estimates of the sensitivity in pixels where the magnitude of the corresponding image is sufficiently high for an accurate estimate.}
  \label{fig:knee-sens}
\end{figure}

Figure \ref{fig:knee-sens} illuminates why this happens based on the understanding presented in Sec. \ref{sec:Theory}---for any specific horizontal location, the coil sensitivity functions as a function of vertical location are approximately scaled versions of each other. When undersampled by every other row, GRAPPA and SPIRiT will attempt to interpolate unknown values from collected data that lies above and below in the Fourier domain. With relatively little spatial variation in the coil sensitivities in those directions, there does not exist a linear combination of coil sensitivities such that they approximate a complex exponential and the interpolation coefficients do not yield an accurate estimate.

\begin{figure}[ht]
  \centering
  \includegraphics[width=3in]{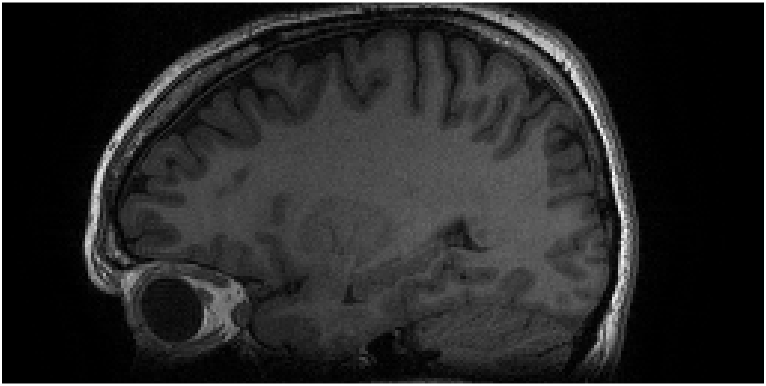}
  \includegraphics[width=3in]{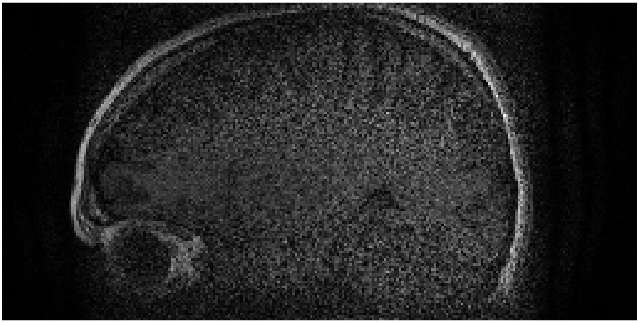}
  \caption{ SPIRiT reconstructions of a sagittal slice of a brain from data collected with an 8-channel birdcage coil using a horizontal (left) and vertical (right) undersampling mask. Undersampling factor of 2, $31\times 31$ ACR, $3\times 3$ kernel.}
  \label{fig:fig3}
\end{figure}

This data was collected with an 8-channel birdcage coil.  Therefore, each coil extends from the most inferior to the most superior portions of the image.  Thus, in the SI (vertical) direction, there is not enough variation to approximate a complex exponential well. This same phenomenon happens when imaging other anatomy with a similar coil arrangement; e.g., imaging the brain with a birdcage coil, as shown in Fig. \ref{fig:fig3}.

\begin{figure}[ht]
  \centering
  \includegraphics[width=5.5in]{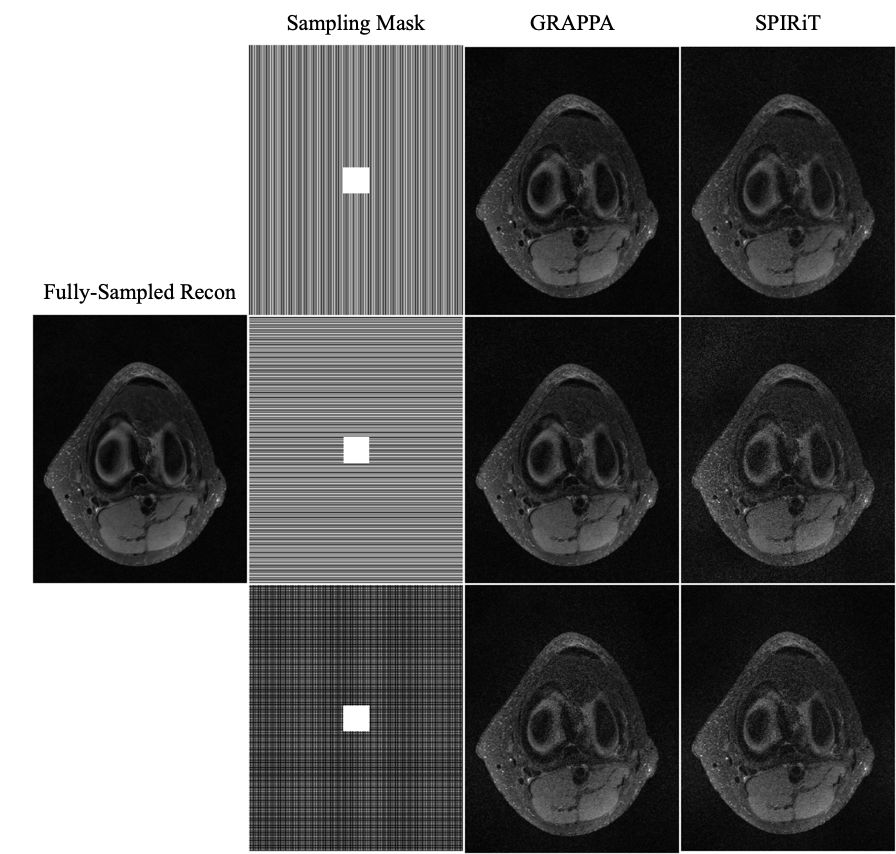}
  \caption{\label{fig:knee_axial_all} Axial slices of the knee.  On the far left is the reconstruction from fully sampled data.  The center-left column is the sampling mask (white are collected samples). The center-right column is the reconstruction from GRAPPA. The right column is the reconstruction with SPIRiT. The top and middle rows have undersampling factors of 2 in a single direction (horizontal and vertical, respectively). The bottom row has a reduction factor of two in both directions. All reconstructions from GRAPPA and SPIRiT were made with a $313\times 31$ ACR and a $3\times 3$ kernel.}
\end{figure}

Figure \ref{fig:knee_axial_all} shows MR images of an axial slice of the knee; this is the same dataset as Fig. \ref{fig:fig1}.  In this case, the quality of the reconstruction is independent of the undersampling direction.  Moreover, when simultaneously undersampling in both directions, the quality remains high; though the signal-to-noise ratio has been reduced due to the reduced scan time \cite{macovski1996noise,nishimura2010principles}. Good quality reconstructions in both directions indicates that the corresponding optimization problems to find the interpolation coefficients for GRAPPA and SPIRiT found good solutions. Owing to our understanding that the optimization problem is attempting to find linear coefficients for the coil sensitivities such that they linearly combine to a weighted complex exponential, the high quality indicates that this is true for at least one direction identified with the GRAPPA and SPIRiT kernels.

\begin{figure}[ht]
  \centering
  \includegraphics[width=2in]{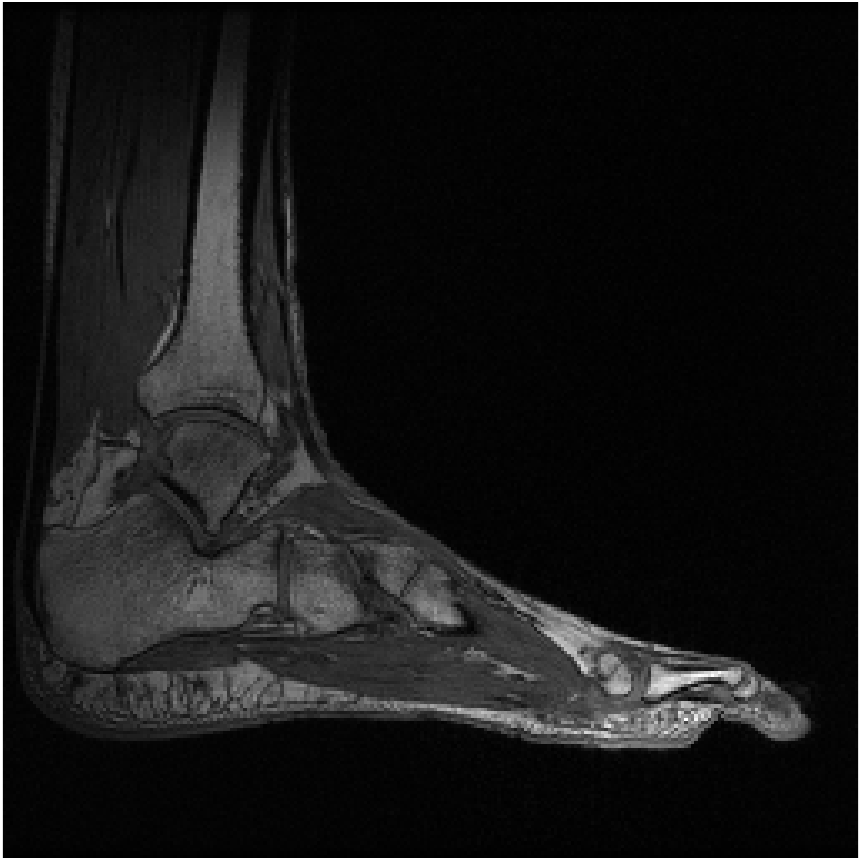}\includegraphics[width=2in]{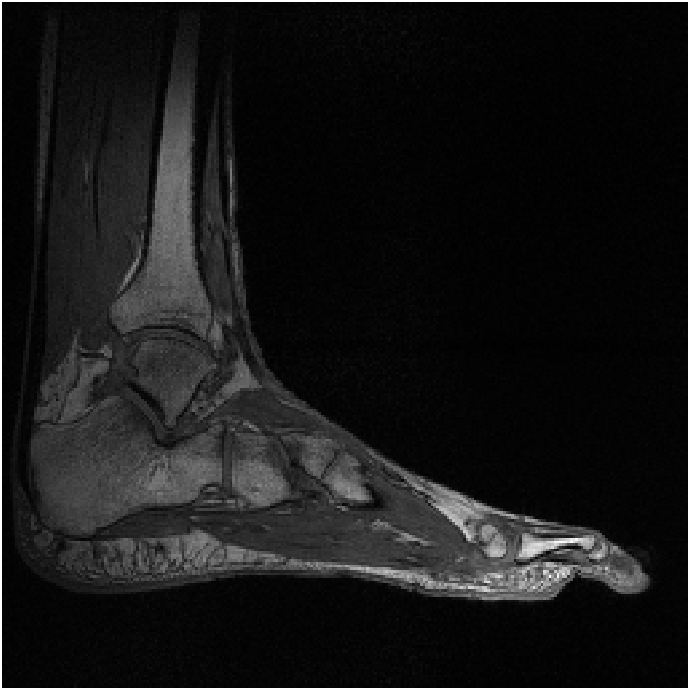} \\
  \includegraphics[width=2in]{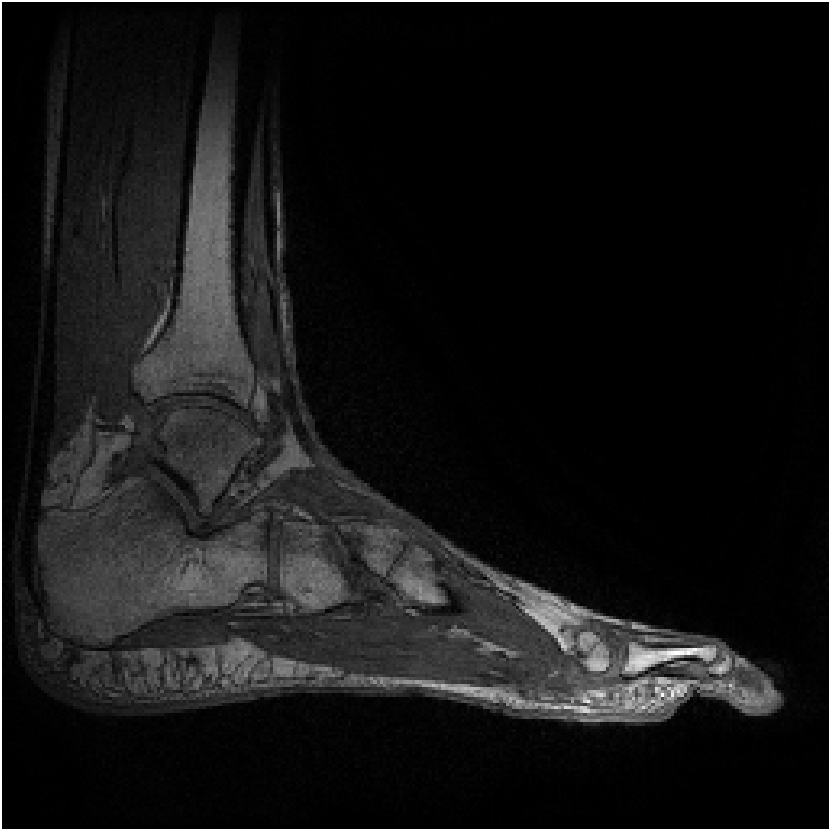}\includegraphics[width=2in]{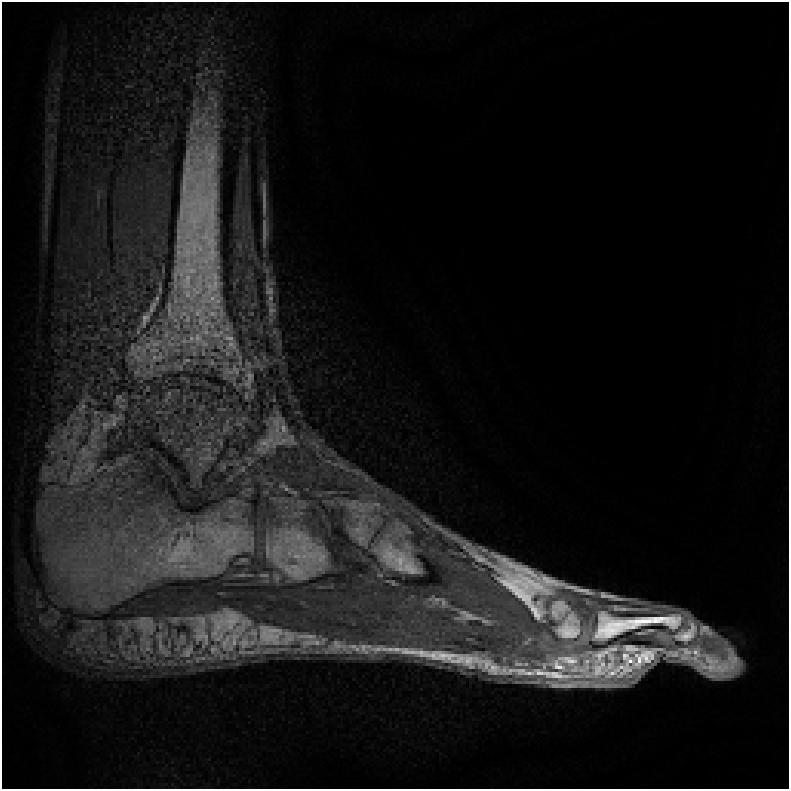} 
  \caption{SPIRiT reconstructions of retrospectively downsampled ankle data. Left: horizontal undersampling pattern. Right: vertical undersampling pattern. Top row: reduction factor of 2. Bottom row: reduction factor of 3. All reconstructions with a $31\times 31$ ACR and a $3\times 3$ kernel.}
  \label{fig:ankleRecons}
\end{figure}

\begin{figure}[ht]
  \centering
  \includegraphics[width=6in]{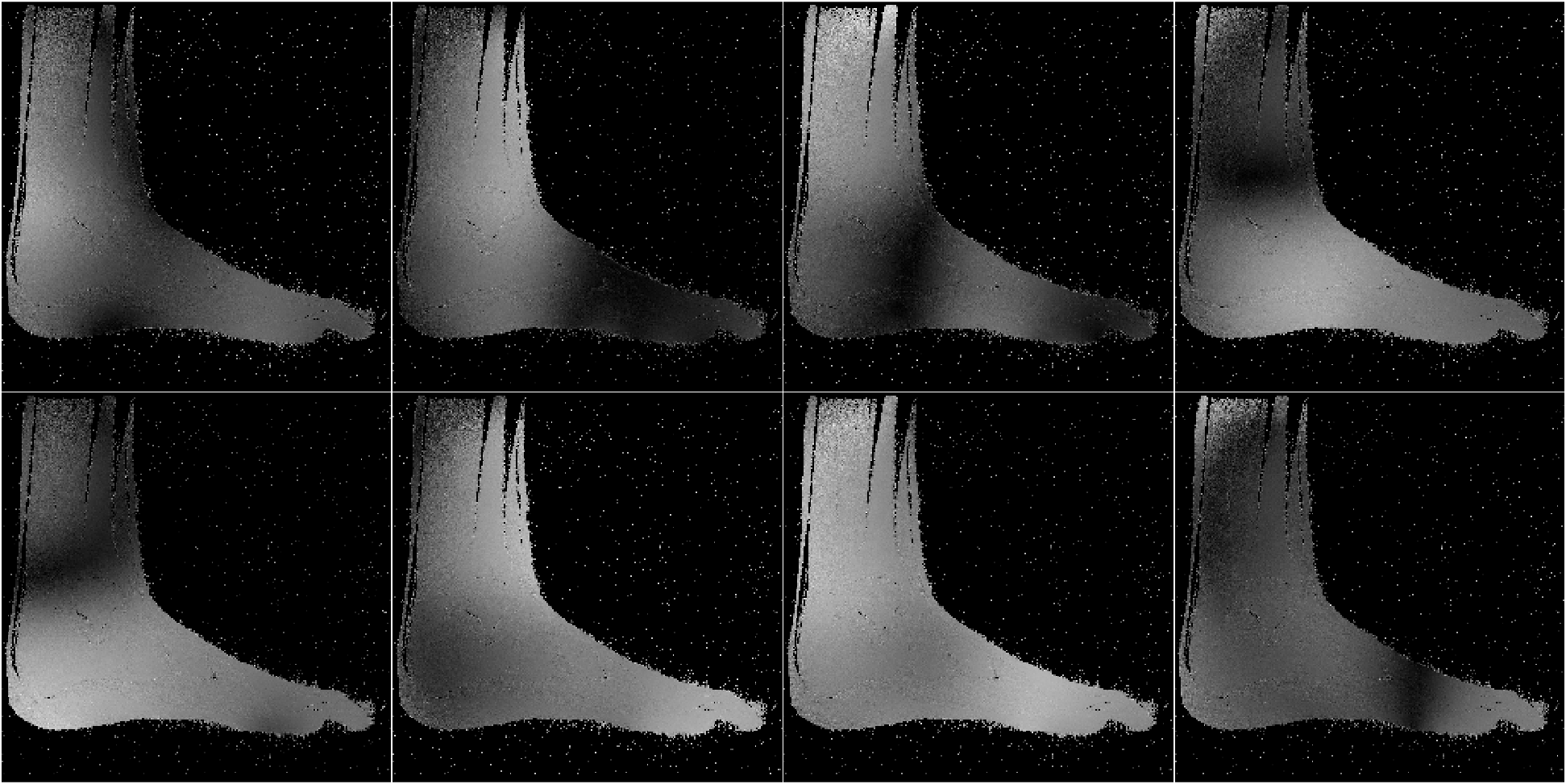}
  \caption{Approximations to the individual sensitivity maps for the data of Fig. \ref{fig:ankleRecons}.  Note that there are only estimates of the sensitivity in pixels where the magnitude of the corresponding image is sufficiently high for an accurate estimate.}
  \label{fig:ankle_sense_maps}
\end{figure}

Figure \ref{fig:ankleRecons} shows reconstructions of a sagittal slice of an ankle from data collected with an 8-channel dedicated ankle coil arrangement. The SPIRiT reconstructions of retrospectively downsampled data results in a high-quality reconstruction independent of the direction of undersampling. Fig. \ref{fig:ankle_sense_maps} shows the sensitivities of each coil. In contrast to the sensitivity maps of Fig. \ref{fig:knee-sens}, each coil shows good spatial variation in both directions.

We varied reconstruction parameters for GRAPPA and SPIRiT reconstructions of the knee and brain to include kernel sizes of $5\times 5$ and $7\times 7$ and to include an undersampling factor of approximately $3$.  In all cases, the same trend was observed: the quality of the reconstruction is highly dependent on the undersampling direction.  (Results are not shown.)

Table \ref{table:metric} shows the horizontal and vertical accuracy metrics for the data studied in this manuscript.  The relative errors for those undersampling directions that yielded a poor quality image are larger than those that yielded a high-quality image.  The axial slice of the knee, which yielded a high-quality reconstruction independent of the undersampling direction, has a small relative error in both directions.  For this dataset, a threshold on the relative error of $0.4$ would identify directions that would yield a poor-quality image.  This, however, is too small a dataset for us to make that a conclusion.  Instead, we present this as a preliminary result and hope to pursue it further in the future. 

\begin{table}
  \centering
  \begin{tabular}{|l|c|c|c|}
    \hline
    Data & Kernel & Vertical Error & Horizontal Error \\
    \hline
    Knee---sagittal slice & 3x3 & 55.0\% (large) & 17.1\% (small) \\ [0.25ex] \hline
    Knee---axial slice & 3x3 &25.2\% (small) & 26.6\% (small) \\ [0.25ex] \hline 
    Ankle---sagittal slice & 3x3 &8.1\% (small) & 5.0\% (small) \\ [0.25ex] \hline
    Brain---sagittal slice & 3x3 & 61.7\% (large) & 5.0\% (small) \\ [0.25ex]
    \hline
    Brain---sagittal slice & 5x5 & 52.6\% (large) & 3.9\% (small) \\ [0.25ex]
    \hline
    Brain---sagittal slice & 7x7 & 47.3\% (large) & 3.6\% (small) \\
    [0.5ex] \hline
  \end{tabular}
  \caption{Examples of the error metric in section \ref{sec:metric} applied to specific cases. The relative errors are large in undersampling directions that lead to poor quality images, and small in undersampling directions that lead to high quality images.}
  \label{table:metric}
\end{table}

\section{Discussion}
We have presented a rigorous physical reasoning that explains why the direction of undersampling can impact reconstruction quality.

Having the requisite spatial variation in the coil sensitivities is a \textit{sufficient} condition for quality reconstruction based on linear predictability.  Other methods of parallel imaging (e.g. SENSE \cite{pruessmann_1999_sense} and model based reconstruction \cite{fessler2010model}), which are not based on linear predictability of the coil sensitivities, may not display such a dependence on the undersampling pattern.  There are other sufficient conditions that may be satisfied for linear predictability which may preclude this difficulty of GRAPPA and SPIRiT.  One such sufficient condition is that the support of the imaged object be less than the field-of-view of the image \cite{haldar_predictability}.  Note further that this condition of support is met by the data studied in this manuscript; however, the quality remains dependent on the direction of undersampling.  This indicates that, for the data studied, the more important consideration is the direction in which the coil sensitivities can be linearly combined to approximate a complex exponential.  

Furthermore, as detailed in Eq. \ref{eqn:auto-smash} of Appendix \ref{app:autosmash}, the linear combination of the sensitivities only needs to approximate a complex exponential over the support of the image.  Generally, the support of the image is a strict subset of the field of view.  Consider the reconstruction of the ankle; there is little coil sensitivity in the superior-anterior quadrant of the image.  However, there is plenty of coil variation over the other quadrants, which is where the ankle and leg are located.  This is why the reconstruction of the ankle is robust regardless of the direction of undersampling. The coils are placed along the ankle and leg and oriented in similar directions (the normal vectors of the coils are all approximately parallel), allowing for local approximations of the sensitivities to complex exponentials.

Results indicate that the accuracy metric can be used to determine undersampling directions that will yield a poor-quality reconstruction, so that those undersampling directions can be avoided during the scan.  In the future, we hope to test this metric on a much larger dataset to ensure that it is reliable. If so, we hope to create an adaptive algorithm that uses the auto-calibration region to identify good undersampling directions prior to collecting the outer portion of the Fourier domain.  By doing so, parallel imaging with linear predictability would become robust to an inappropriately selected undersampling direction.

Existing methods for combining parallel imaging with compressed sensing often utilize a variable density poisson disc sampling pattern \cite{dwork2021fast} based on the intuition that samples should not be placed too close together since parallel imaging permits interpolation between samples \cite{vasanawala2011practical}.  Based on the work of this manuscript, parallel imaging with linear predictability may not be able to accurate interpolate between samples that are spaced in a given direction.  If the parameter that governs variable density for the sampling pattern is radially symmetric, this indicates that the reconstruction relies more heavily on the sparsity assumptions of compressed sensing in one direction over another and suggests that there may be a sampling pattern that yields a higher quality when combining parallel imaging with compressed sensing.

\section{Conclusion}
We showed that the quality of parallel imaging reconstruction algorithms based on linear predictability, such as GRAPPA and SPIRiT, has a dependence on the direction of undersampling. This directional dependence is related to the amount of spatial variance in the individual coil sensitivities.  The impact to the quality of reconstructions is the difference between having a diagnostic image or not.  We note that this happens for specific coil configurations and body slices, whereas other configurations are robust to this problem.

\appendix
\section{AUTO-SMASH}
\label{app:autosmash}

In a coil array with $J$ elements, the $j^\text{th}$ coil has a distinct sensitivity function $C_j:\mathbb{R}^2\rightarrow\mathbb{C}$.  A composite sensitivity is generated as a linear combination of individual coil sensitivities with linear coefficients $n_j^{(0)}$ as follows:
\begin{equation}
C_0^{\text{comp}}(x, y) = \sum_{j=1}^{J}n_j^{(0)}C_j(x, y).
\end{equation}

The composite two-dimensional MR signal takes the form,
\begin{align} \label{eqn:comp_mr}
S^{\text{comp}}(k_x, k_y) &= \sum_{j=1}^Jn_j^{(0)}S_j(k_x, k_y) \nonumber\\
&= \iint dx \, dy \, \sum_{j=1}^Jn_j^{(0)}C_j(x, y)\rho(x, y)\text{exp}\{-ik_xx - ik_yy\}\nonumber \\
&= \iint dx \, dy \, C_0^{\text{comp}}(x, y)\rho(x, y)\exp{\{-ik_xx - ik_yy\}} = \mathcal{F}\{C_0^{(\text{comp})}\rho\}(k_x, k_y),
\end{align}
the Fourier transformation of $C_0^{\text{comp}}\rho$.

Suppose that there is another set of complex weights $\{n_j^{(m)}\}$ such that the linear combination of coil sensitivities yields the following composite sensitivity:
\begin{equation}
\label{eqn:smash_weights}
    C_m^{\text{comp}}(x, y) = \sum_{j=1}^Jn_j^{(m)}\,C_j(x, y) \approx C_0^{\text{comp}} \exp\left(i\,m\,\Delta k_y\, y\right).
\end{equation}
Importantly, this approximation only needs to be valid over the support of the image. With these linear coefficients, the composite MR signal becomes
\begin{equation}
  \begin{aligned}\label{eqn:auto-smash}
    \sum_{j=1}^Jn_j^{(m)}S_j(k_x, k_y) &= \sum_{j=1}^Jn_j^{(m)}\iint_{-\infty}^{\infty} dx\,dy\,C_j(x, y)\rho(x, y)\exp\{-ik_xx -i k_yy\} \\
    &= \iint_\Omega dx\,dy \left[\sum_{j=1}^Jn_j^{(m)}\,C_j(x, y)\right]\rho(x, y)\exp\{-i\,k_x\,x -i\, k_y\,y\} \\
    &\approx \iint_\Omega dx\,dy \;C_0^{\text{comp}}\exp\left(i\,m\Delta k_y y\right)\,\rho(x, y)\exp\left(-i\,k_x\,x -i\,k_y\,y\right) \\
    &= \iint_{-\infty}^{\infty} dx\,dy \;C_0^{\text{comp}}\,\rho(x, y)\exp\left(-i\,k_x\,x -i\,(k_y - m\Delta k_y)\,y\right), \\
    &=\mathcal{F}\{C_0^{\text{comp}}\rho\}(k_x, k_y - m\Delta k_y),
  \end{aligned}
\end{equation}
where $\Omega$ is the support of $\rho$.
The $n^{(m)}$ coefficients serve to interpolate Fourier values at a distance of $m\Delta k_y$.  Note that for the special case where $C_0^{\text{comp}}\approx 1$, the Fourier coefficients are those of $\rho$.

The innovation of AUTO-SMASH is to use collected auto-calibration signal (ACS) lines to estimate the weights $\{n_j^{(m)}\}$ for Eq. \ref{eqn:smash_weights}. These ACS data, $S_j^{ACR}$ are exactly shifted by the amount $m \Delta k_y$. The composite signal generated using weights $\{n_j^{(0)}\}$ according to Eq. \ref{eqn:comp_mr} yields
\begin{equation}
  \label{eqn:acs1}
  S^{\text{comp}}(k_x, k_y - m\Delta k_y) = \sum_{j=1}^Jn_j^{(0)}S_j^{ACS}(k_x, k_y - m\Delta k_y).
\end{equation}

Alternatively, following Eq. \ref{eqn:auto-smash}, we can write
\begin{equation}
  \label{eqn:acs2}
  S^{\text{comp}}(k_x, k_y - m\Delta k_y) = \sum_{j=1}^{J}n_j^{(m)}S_j(k_x, k_y).
\end{equation}

Equating Eq. \ref{eqn:acs1} and Eq. \ref{eqn:acs2} yields
\begin{equation} \label{eqn:as-summations}
\sum_{j=1}^{J}n_j^{(m)}S_j(k_x, k_y) = \sum_{j=1}^Jn_j^{(0)}S_j^{ACS}(k_x, k_y - m\Delta k_y)
\end{equation}

Write the right-hand side of Eq. \ref{eqn:as-summations} simply as $S^{\text{comp}}(k_x, k_y - m\Delta k_y)$ to reinforce that is the final, combined image produced using the original weights. For each $k_x$ this is a (complex) scalar, and the left-hand side is a linear combination of the collected MR signals. We are attempting to solve for $n_j^{(m)}$.

To write this as a linear system, denote $\Sigma$, $n^{(m)}$ and $b$ as follows:
\begin{equation}\nonumber
\Sigma = 
\underbrace{\begin{bmatrix}
S_1(k_{x_1}, k_y) & S_2(k_{x_1}, k_y) & \ldots & S_J(k_{x_1}, k_y) \\
S_1(k_{x_2}, k_y) & S_2(k_{x_2}, k_y) & \ldots & S_J(k_{x_2}, k_y) \\
\vdots & \ddots &&\vdots \\
S_1(k_{x_{n_x}}, k_y) & S_2(k_{x_{n_x}}, k_y) & \ldots & S_J(k_{x_{n_x}}, k_y)
\end{bmatrix},}_{n_x \times J } \hspace{0.4em}
n^{(m)} = \underbrace{\begin{bmatrix}n_1^{(m)} \\ n_2^{(m)} \\ \ldots \\ n_J^{(m)} \end{bmatrix},}_{J \times 1} \hspace{0.4em}
b = \underbrace{\begin{bmatrix}S^{\text{comp}}(k_{x_1}, k_y - m\Delta k_y) \\ S^{\text{comp}}(k_{x_2}, k_y - m\Delta k_y) \\ \ldots \\ S^{\text{comp}}(k_{x_{n_x}}, k_y - m\Delta k_y) \end{bmatrix}.}_{n_x \times 1}
\end{equation}
To find the interpolation coefficients $n^{(m)}$, one can minimize $\|\Sigma\,n^{(m)}-b\|_2$.

Once the weights $n^{(m)}$ are determined, the matrix-vector multiplication $\Sigma\,n^{(m)}$ for $\Sigma$ constructed for a specific $(k_x,k_y)$ will estimate the \textit{composite} Fourier value at $(k_x,k_y - m\,\Delta k_y)$.


\begin{thebibliography}{10}

\bibitem{deshmane2012parallel}
Anagha Deshmane, Vikas Gulani, Mark~A Griswold, and Nicole Seiberlich.
\newblock Parallel {MR} imaging.
\newblock {\em Journal of Magnetic Resonance Imaging}, 36(1):55--72, 2012.

\bibitem{lustig2007sparse}
Michael Lustig, David Donoho, and John~M Pauly.
\newblock Sparse {MRI}: The application of compressed sensing for rapid mr
  imaging.
\newblock {\em Magnetic Resonance in Medicine}, 58(6):1182--1195, 2007.

\bibitem{baron2018rapid}
Corey~A Baron, Nicholas Dwork, John~M Pauly, and Dwight~G Nishimura.
\newblock Rapid compressed sensing reconstruction of {3D} non-cartesian {MRI}.
\newblock {\em Magnetic resonance in medicine}, 79(5):2685--2692, 2018.

\bibitem{dwork2021utilizing}
Nicholas Dwork, Daniel O’Connor, Corey~A. Baron, Ethan~M.I. Johnson, Adam~B
  Kerr, John~M. Pauly, and Peder~E.Z. Larson.
\newblock Utilizing the wavelet transform’s structure in compressed sensing.
\newblock {\em Signal, Image and Video Processing}, 15(7):1407--1414, 2021.

\bibitem{dwork2022utilizing}
Nicholas Dwork and Peder~EZ Larson.
\newblock Utilizing the structure of a redundant dictionary comprised of
  wavelets and curvelets with compressed sensing.
\newblock {\em Journal of Electronic Imaging}, 31(6):063043, 2022.

\bibitem{noll1991homodyne}
Douglas~C Noll, Dwight~G Nishimura, and Albert Macovski.
\newblock Homodyne detection in magnetic resonance imaging.
\newblock {\em Transactions on Medical Imaging}, 10(2):154--163, 1991.

\bibitem{sandino2020compressed}
Christopher~M Sandino, Joseph~Y Cheng, Feiyu Chen, Morteza Mardani, John~M
  Pauly, and Shreyas~S Vasanawala.
\newblock Compressed sensing: From research to clinical practice with deep
  neural networks: Shortening scan times for magnetic resonance imaging.
\newblock {\em Signal Processing Magazine}, 37(1):117--127, 2020.

\bibitem{hammernik2019sigma}
Kerstin Hammernik, Jo~Schlemper, Chen Qin, Jinming Duan, Ronald~M Summers, and
  Daniel Rueckert.
\newblock $\sigma$-net: Systematic evaluation of iterative deep neural networks
  for fast parallel mr image reconstruction.
\newblock {\em arXiv}, 1912.09278, 2019.

\bibitem{knoll2020advancing}
Florian Knoll, Tullie Murrell, Anuroop Sriram, Nafissa Yakubova, Jure Zbontar,
  Michael Rabbat, Aaron Defazio, Matthew~J Muckley, Daniel~K Sodickson,
  C~Lawrence Zitnick, et~al.
\newblock Advancing machine learning for {MR} image reconstruction with an open
  competition: Overview of the 2019 {fastMRI} challenge.
\newblock {\em Magnetic resonance in medicine}, 84(6):3054--3070, 2020.

\bibitem{jakob1998auto}
Peter~M Jakob, Mark~A Grisowld, Robert~R Edelman, and Daniel~K Sodickson.
\newblock {AUTO-SMASH}: a self-calibrating technique for {SMASH} imaging.
\newblock {\em Magnetic Resonance Materials in Physics, Biology and Medicine},
  7(1):42--54, 1998.

\bibitem{griswold2002generalized}
Mark~A Griswold, Peter~M Jakob, Robin~M Heidemann, Mathias Nittka, Vladimir
  Jellus, Jianmin Wang, Berthold Kiefer, and Axel Haase.
\newblock Generalized autocalibrating partially parallel acquisitions
  ({GRAPPA}).
\newblock {\em Magnetic Resonance in Medicine}, 47(6):1202--1210, 2002.

\bibitem{lustig2010spirit}
Michael Lustig and John~M. Pauly.
\newblock {SPIRiT}: Iterative self-consistent parallel imaging reconstruction
  from arbitrary k-space.
\newblock {\em Magnetic Resonance in Medicine}, 64(2):457--471, 2010.

\bibitem{haldar_predictability}
Justin~P. Haldar and Kawin Setsompop.
\newblock Linear predictability in magnetic resonance imaging reconstruction:
  Leveraging shift-invariant fourier structure for faster and better imaging.
\newblock {\em Signal Processing Magazine}, 37(1):69--82, 2020.

\bibitem{seiberlich2007non}
Nicole Seiberlich, Felix~A Breuer, Martin Blaimer, Kestutis Barkauskas, Peter~M
  Jakob, and Mark~A Griswold.
\newblock Non-cartesian data reconstruction using {GRAPPA} operator gridding
  ({GROG}).
\newblock {\em Magnetic Resonance in Medicine: An Official Journal of the
  International Society for Magnetic Resonance in Medicine}, 58(6):1257--1265,
  2007.

\bibitem{wright2014non}
Katherine~L Wright, Jesse~I Hamilton, Mark~A Griswold, Vikas Gulani, and Nicole
  Seiberlich.
\newblock Non-{Cartesian} parallel imaging reconstruction.
\newblock {\em Journal of Magnetic Resonance Imaging}, 40(5):1022--1040, 2014.

\bibitem{luo2019grappa}
Tianrui Luo, Douglas~C Noll, Jeffrey~A Fessler, and Jon-Fredrik Nielsen.
\newblock A {GRAPPA} algorithm for arbitrary {2D/3D} non-cartesian sampling
  trajectories with rapid calibration.
\newblock {\em Magnetic resonance in medicine}, 82(3):1101--1112, 2019.

\bibitem{beatty2007method}
PJ~Beatty, AC~Brau, S~Chang, S~Joshi, CR~Michelich, E~Bayram, TE~Nelson,
  RJ~Herfkens, and JH~Brittain.
\newblock A method for autocalibrating {2D} accelerated volumetric parallel
  imaging with clinically practical reconstruction times.
\newblock In {\em Proceedings of the International Society for Magnetic
  Resonance in Medicine}, volume 1749, 2007.

\bibitem{epstein2007introduction}
Charles~L Epstein.
\newblock {\em Introduction to the mathematics of medical imaging}.
\newblock SIAM, 2007.

\bibitem{roemer1990nmr}
Peter~B Roemer, William~A Edelstein, Cecil~E Hayes, Steven~P Souza, and
  Otward~M Mueller.
\newblock The {NMR} phased array.
\newblock {\em Magnetic resonance in medicine}, 16(2):192--225, 1990.

\bibitem{beck2009fast}
Amir Beck and Marc Teboulle.
\newblock A fast iterative shrinkage-thresholding algorithm for linear inverse
  problems.
\newblock {\em Journal on imaging sciences}, 2(1):183--202, 2009.

\bibitem{ong2018mridata}
Frank Ong, Shahab Amin, Shreyas Vasanawala, and Michael Lustig.
\newblock Mridata.org: An open archive for sharing {MRI} raw data.
\newblock In {\em Proceedings of the International Society of Magnetic
  Resonance in Medicine}, volume~26, 2018.

\bibitem{giovannetti2002fast}
Giulio Giovannetti, Luigi Landini, Maria~Filomena Santarelli, and Vincenzo
  Positano.
\newblock A fast and accurate simulator for the design of birdcage coils in
  {MRI}.
\newblock {\em Magnetic Resonance Materials in Physics, Biology and Medicine},
  15(1):36--44, 2002.

\bibitem{macovski1996noise}
Albert Macovski.
\newblock Noise in {MRI}.
\newblock {\em Magnetic resonance in medicine}, 36(3):494--497, 1996.

\bibitem{nishimura2010principles}
Dwight~G Nishimura.
\newblock {\em Principles of magnetic resonance imaging}.
\newblock Standford Univ., 2010.

\bibitem{pruessmann_1999_sense}
Klaas~P. Pruessmann, Markus Weiger, Markus~B. Scheidegger, and Peter Boesiger.
\newblock Sense: Sensitivity encoding for fast {MRI}.
\newblock {\em Magnetic Resonance in Medicine}, 42(5):952–962, Nov 1999.

\bibitem{fessler2010model}
Jeffrey~A Fessler.
\newblock Model-based image reconstruction for {MRI}.
\newblock {\em Signal processing magazine}, 27(4):81--89, 2010.

\bibitem{dwork2021fast}
Nicholas Dwork, Corey~A Baron, Ethan~MI Johnson, Daniel O'Connor, John~M Pauly,
  and Peder~EZ Larson.
\newblock Fast variable density poisson-disc sample generation with directional
  variation for compressed sensing in {MRI}.
\newblock {\em Magnetic Resonance Imaging}, 77:186--193, 2021.

\bibitem{vasanawala2011practical}
SS~Vasanawala, MJ~Murphy, Marcus~T Alley, P~Lai, Kurt Keutzer, John~M Pauly,
  and Michael Lustig.
\newblock Practical parallel imaging compressed sensing {MRI}: Summary of two
  years of experience in accelerating body {MRI} of pediatric patients.
\newblock In {\em international symposium on biomedical imaging: From nano to
  macro}, pages 1039--1043. {IEEE}, 2011.

\end{thebibliography}

\end{document}